# Hybrid integrated ultra-low linewidth coil stabilized isolator-free widely tunable external cavity laser

David A. S. Heim[1], Debapam Bose[1], Kaikai Liu[1], Andrei Isichenko[1], and Daniel J. Blumenthal[1*]

[1]Department of Electrical and Computer Engineering, University of California Santa Barbara, Santa Barbara, CA 93106 USA.
[*]Corresponding author (danb@ucsb.edu)

## ABSTRACT

Precision applications including quantum computing and sensing, mmWave/RF generation, and metrology, demand widely tunable, ultra-low phase noise lasers. Today, these experiments employ table-scale systems with bulk-optics and isolators to achieve requisite noise, stability, and tunability. Photonic integration will enable scalable, reliable and portable solutions. Here we report a hybrid-integrated external cavity widely tunable laser stabilized to a 10-meter-long integrated coil-resonator, achieving record-low 3 – 7 Hz fundamental linewidth across a 60 nm tuning range and 27 – 60 Hz integral linewidth with 1.8E-13 Allan deviation at 6.4 ms across 40 nm, delivering orders of magnitude frequency noise and integral linewidth reduction over state of the art. Stabilization is achieved without an optical isolator, leveraging resilience to optical feedback of 30 dB beyond that of commercial DFB lasers. The laser and reference cavity are fabricated in the same Si3N4 CMOS-compatible process, unlocking a path towards fully integrated visible to ShortWave-IR frequency-stabilized lasers.

# INTRODUCTION

Ultra-narrow linewidth widely tunable stabilized lasers are critical for a range of precision applications including optical atomic clocks[1,2], quantum computing[3-6], metrology[7,8], quantum and fiber sensing[9-11], and low phase noise mmWave and RF generation[12,13]. Of paramount importance to these applications is the phase noise as measured from low to high carrier offset frequencies and characterized in part by the instantaneous and integral linewidths. To achieve ultra-low linewidths, these lasers systems utilize large mode volume lab-scale external cavity lasers, bulk-optic reference cavities[14,15] and optical isolation in the laser stabilization circuit. Integration using a wavelength transparent platform will improve reliability, reduce size, weight, power, and cost, and enable scalability, portability, and systems-on-chip solutions across the visible to ShortWave-IR (SWIR). Yet, to date, integration of widely tunable, cavity stabilized lasers, in a CMOS foundry compatible integrated platform, has remained elusive. Low instantaneous linewidth integrated lasers include self-injection locked (SIL)[16-19], stimulated Brillouin scattering (SBS)[20-22], external cavity DBR[23], and external cavity lasers (ECLs)[24,25]. In particular, the external cavity tunable laser (ECTL) design[26-28] is used due to its wide wavelength tuning range, low fundamental linewidth, and ability to stabilize the laser output to an external optical reference cavity for integral linewidth reduction and carrier stabilization. Photonic integration of stabilized ECTLs is a critical step forward for robust solutions, operating from the visible to SWIR, to serve as stand-alone sources and as pumps for optical frequency combs, nonlinear wavelength conversion, and Brillouin lasers.

Silicon nitride ($Si_3N_4$) is an ideal integration platform[29] due to its low propagation loss that extends from the visible to SWIR. The combination of ultra-high Q resonators[30,31], the ability to integrate gain media[32], and waveguide-compatible control and modulation[33-35] enable a wide range of systems-on-chip solutions. Silicon nitride photonics have been used to realize narrowly tunable low fundamental linewidth lasers[36-39] and large mode-volume resonator reference cavities[40,41]. The high Q resonators enable increased intra-cavity photon lifetime and photon number needed to reduce the fundamental linewidths, while large mode-volume laser and reference cavities decrease the intrinsic thermorefractive noise (TRN)[42]. Additionally, a high laser resonator Q and other techniques can improve the resilience to optical feedback[43-47]. Hybrid integration of $Si_3N_4$ ECLs[23,25,26,28,36,37,39] is an effective way to combine the benefits of ultra-low loss waveguides and high Q resonators with III-V semiconductor gain materials at a wide range of wavelengths. However, achieving both low fundamental and low integral linewidths across widely tunable wavelength ranges, and without the need for optical isolation in a common integration platform that supports the ECTL, reference cavity, and other photonic components has not been achieved.

Here we present a significant advance in chip-scale widely tunable stabilized laser technology by demonstrating a coil stabilized isolator-free hybrid integrated $Si_3N_4$ ECTL with a 60 nm tuning range, fundamental linewidth (FLW) of 3 to 7 Hz across a full 60 nm tuning range and 27 - 60 Hz $1/\pi$ integral linewidth (ILW) across a 40 nm tuning range. These results represent a frequency noise reduction of over 6-orders of magnitude, almost 2-orders of magnitude reduction in integral linewidth over the free running linewidth, and 65 dB side mode suppression ratio (SMSR) across the tuning range. Linewidth reduction and carrier stabilization are achieved by directly locking the ECTL, without optical isolation, to a silicon nitride 10-meter long coil resonator reference. The 20 MHz free-spectral range (FSR) of the 10-meter coil enables stabilization at almost any wavelength across the 60 nm tuning range. We also measure an Allan deviation (ADEV) of $1.8 \times 10^{-13}$ at 6.4 ms and 5.0 kHz/s drift. An accurate ILW measurement is achieved using an ultra-low expansion (ULE) cavity-stabilized frequency comb to heterodyne beatnote measure the laser noise down to 1 Hz frequency offset across a 40 nm tuning range. We demonstrate that the feedback resilience of the 3.5 million intrinsic-Q Vernier rings in the ECTL provides an inherent isolation of ~30 dB relative to a typical commercial III-V DFB laser. The ECTL and 10-meter coil-reference cavity are fabricated in the same 80 nm thick silicon nitride low loss platform. This common design and fabrication process combined with the feedback resilience of the ECTL enables a path towards full integration of widely tunable, narrow-linewidth, frequency stabilized lasers at a wide range of quantum, atomic, fiber communication, and other wavelengths from the visible through the SWIR.

## RESULTS

**Coil stabilized ECTL architecture and design.** The coil stabilized integrated tunable laser experiment, shown in Fig. 1a, consists of a widely tunable silicon nitride ECTL (left side of Fig. 1a) directly connected through an electrooptic modulator (EOM) for sideband locking to a 10-meter coil resonator reference cavity (right side of Fig. 1a). The coil resonator output is tapped and photodetected and fed-back to control the ECTL lasing frequency using a Pound-Drever-Hall (PDH) circuit. The ECTL and coil resonator are fabricated in the same 80 nm thick silicon nitride fabrication process and utilize the fundamental TE mode where the waveguide width can be adjusted to optimize the tradeoff between bend radius and propagation losses due to sidewall scattering (see Supplementary Note 2 & 3 for more details on waveguide design).

The hybrid integrated ECTL (Fig. 1c) consists of a silicon nitride waveguide section with two actuatable high-Q intra-cavity ring resonators, an adjustable phase section, and a tunable Sagnac loop mirror and a gain section consisting of a butt-coupled InP reflective semiconductor optical amplifier (RSOA). The two high-Q rings, connected in an add-drop configuration, increase the intra-cavity photon lifetime and provide instantaneous linewidth reduction. Thermo-optically tuned actuators on the ring resonators enable wide single mode tuning utilizing the Vernier effect, and the tunable Sagnac loop mirror serves as a tunable broadband cavity reflector. Additional fine lasing mode tuning is achieved using the thermally actuated phase section. Hybrid integration is achieved by edge coupling a Thorlabs single angled facet C-band reflective SOA (SAF 1128C) to an 18 μm-wide angled waveguide fabricated at the input to the $Si_3N_4$ ECTL chip. The RSOA is mounted on a heat sink and wire bonded to an electronic circuit board. The input waveguide is designed to optimize the modal overlap between the gain chip and the $Si_3N_4$ PIC before tapering down to 2.6 μm wide to achieve ultra-low propagation loss in the external cavity circuit (see Methods). The thermo-optically controlled ring resonators, with radii of 1998.36 and 2002.58 μm, have an FSR of ~126.5 pm and therefore a Vernier FSR of approximately 59.9 nm. The rings have an intrinsic-Q of 3.5 million (see Supplementary Note 2) and are designed to be over-coupled to reduce the lasing threshold, resulting in a loaded-Q of 0.65 million. The laser output waveguide is coupled to a lensed fiber.

The silicon nitride 10-meter coil resonator (Fig. 1e) is a 25 mm x 25 mm chip consisting of a bus coupled coil waveguide geometry (see Supplementary Note 3) with a measured propagation loss of 0.2 dB/m and an intrinsic-Q of approximately 200 million. The 10-meter-long cavity reduces the TRN floor of the stabilized laser due to its large mode volume[42,48]. Additionally, the 10-meter coil length provides a fine 20 MHz FSR with good fringe extinction ratio over more than 80 nm of bandwidth that provides an almost continuous range of lock frequencies across the widely tunable laser range of 60 nm. The frequency noise performance is measured using an unbalanced optical frequency discriminator (OFD) for offset frequencies > 1 kHz and an ultra-low expansion (ULE) cavity-stabilized fiber frequency comb for offset frequencies of 1 kHz down to 1 Hz.

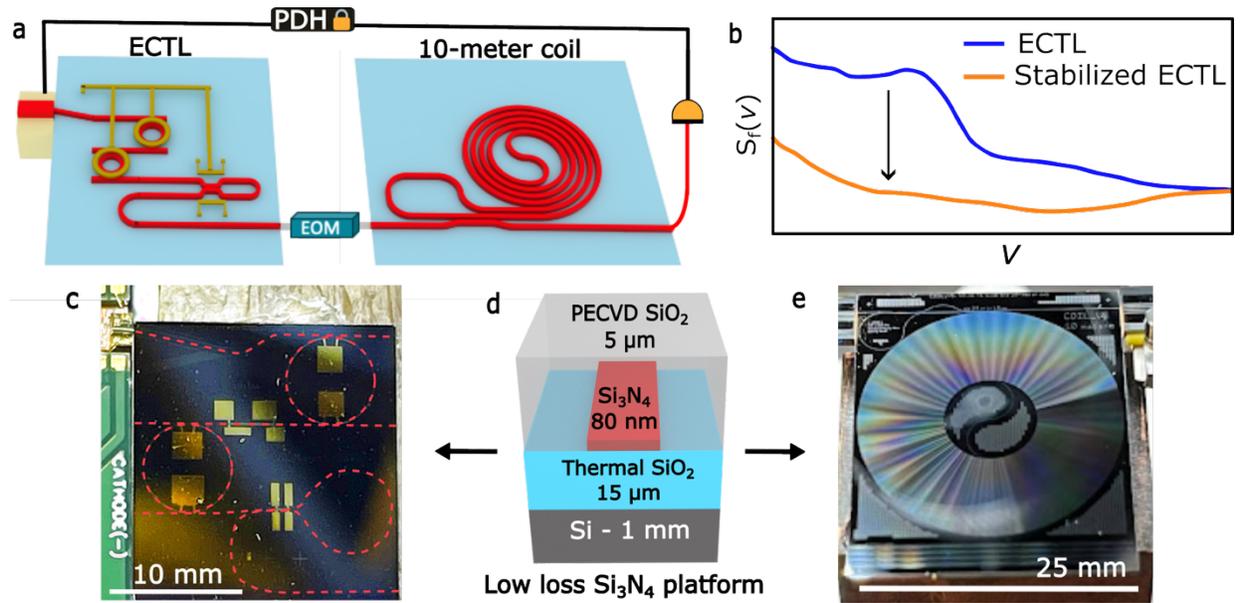

**Fig. 1 Experimental setup of the integrated ECTL stabilized to an integrated coil reference cavity. a** Schematic of the experimental setup where the external cavity tunable laser (ECTL) is PDH-locked to an integrated 10-m coil reference cavity without an optical isolator. Hybrid integration of a reflective semiconductor optical amplifier (RSOA) provides a gain element. The high quality factor (Q) silicon nitride ($Si_3N_4$) rings serve as an external cavity and provide instantaneous linewidth narrowing, and the large mode volume $Si_3N_4$ coil resonator provides a frequency reference for linewidth reduction and laser stabilization. **b** The frequency noise spectral energy (offset from the carrier) of the free-running ECTL ($S_f(v)$ blue) and the noise reduction, particularly of the low-frequency noise components, resulting from stabilizing the ECTL to the 10-meter coil reference cavity (orange). **c** Image of the hybrid-integrated ECTL, the RSOA is visible in the top left, the red-dashed lines highlight the $Si_3N_4$ waveguides, and gold pads indicate the metal heaters. **d** Schematic cross-section of the 80 nm thick $Si_3N_4$ low loss waveguide platform. **e** Image of the 10-m coil resonator.

**ECTL performance.** The ECTL performance is summarized in Figure 2. We demonstrate a 60 nm-wide single-mode operating range, from 1518.5 to 1578 nm (Fig. 2a), with a side-mode suppression ratio (SMSR) of 65 dB (Fig. 2c) across the tuning range. The ECTL ring heater tuning efficiency is 54.5 nm/W per actuator (Fig. 2b). We measure a lasing threshold current at 1550 nm of 63 mA and a fiber-coupled output power ranging from 0.23 mW at 1520 nm to 4.37 mW at 1578 nm (Fig. 2d). The change in output power is primarily due to the wavelength dependence of the evanescent coupler in the Sagnac loop mirror. The frequency noise (FN) spectrum of the free-running widely tunable ECTL without coil-resonator stabilization is shown for 1550 nm emission in Fig. 2e, indicating the fundamental and integral linewidths of the free-running laser at this wavelength. For intermediate offset frequencies, ~ 2 kHz to 200 kHz, the FN closely follows the calculated TRN limit of the ECTL rings (green-dashed curve). The laser noise reaches the white frequency noise (WFN) floor characteristic of the Lorentzian fundamental linewidth (FLW) at > 1 MHz frequency offset. The periodic high frequency spikes are from the fiber Mach-Zehnder interferometer frequency noise measurement setup. Photothermal noise dominates at lower than 2 kHz and the MZI photodiode (PD) noise dominates at frequencies greater than 100 MHz. We measure a WFN floor at 1550 nm of 1.93 $Hz^2$/Hz, corresponding to a 6.08 Hz (light-blue dashed line) FLW. The integral linewidth (ILW) of the free-running ECTL at this wavelength is calculated using the $1/\pi$ reverse integration method[40] to be 1.75 kHz and shown in the purple shaded region under the FN curve. Across the full tuning range, we measure the fundamental linewidth of the free-running ECTL to be in the range of 3 – 7 Hz as summarized in Fig. 2f (see Supplementary Note 4 for the full dataset).

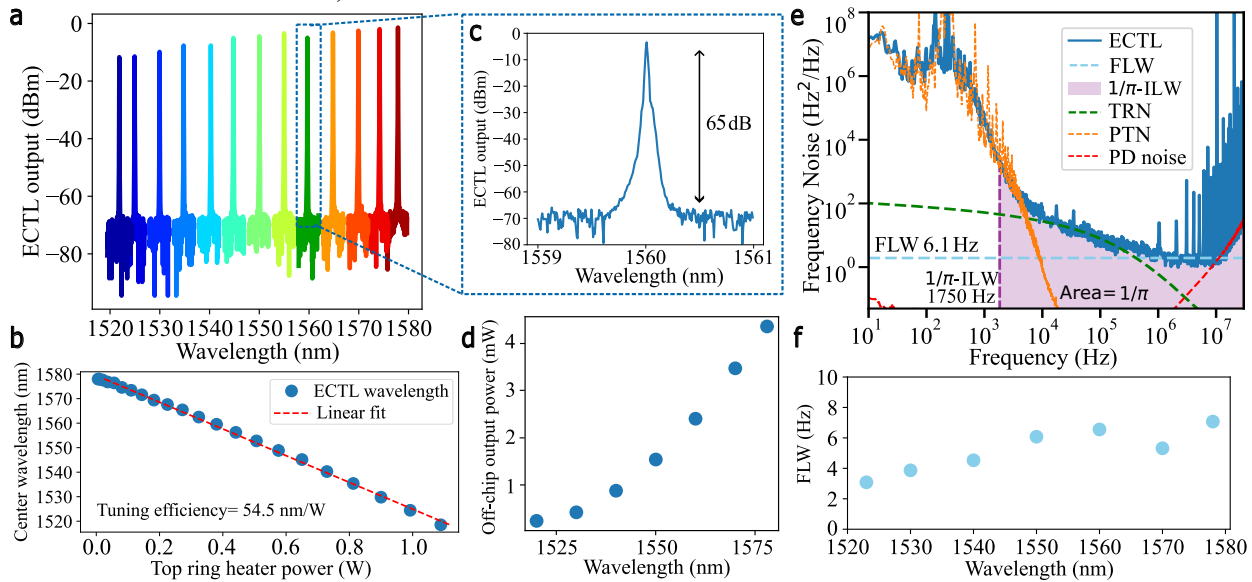

**Fig. 2 External cavity tunable laser performance. a** Single mode laser output from 1520 – 1580 nm. **b** Thermo-optic tuning of the top ring of the external cavity tunable laser (ECTL). **c** ECTL operation at 1560 nm with a measured side mode suppression ratio (SMSR) of ~ 65 dB, measured on an optical spectrum analyser (OSA) with resolution bandwidth (RBW) < 0.01 nm. **d** Fiber-coupled ECTL output power measured across the tuning range. **e** Frequency noise spectrum of the free-running ECTL at 1550 nm (blue, solid) measured using a fiber-Mach Zehnder Interferometer (fiber-MZI) as an optical frequency discriminator (OFD). The blue dashed line plots the measured ECTL fundamental linewidth (FLW) of 6.1 Hz, and the shaded purple region shows the area under the curve that contributes to the 1/π-integral linewidth (ILW) of 1750 Hz. The green and orange dashed curves are simulated estimates of the thermorefractive noise (TRN) and photothermal noise (PTN) limits of the ECTL rings, and the red dashed curve is the OFD photodetector noise. The high frequency spurs at multiples of 1 MHz correspond to the free-spectral range of the OFD fiber-MZI and do not contribute to the integral linewidth calculation. **f** FLW of the ECTL measured across the 60 nm tuning range. Higher reflectivity of the Sagnac loop mirror and weaker ring-bus coupling may contribute to the decrease in FLW at shorter wavelengths.

**Coil stabilized ECTL performance**. We stabilize the ECTL to the 10-meter-long silicon nitride integrated coil reference cavity without an optical isolator between the laser and reference cavity using an EOM to generate locking sidebands of 10 - 20 MHz and a PDH error signal fed back directly to the ECTL gain chip current using a Vescent D2-125 laser servo. The ECTL takes on the FN characteristics of the coil resonator for frequency offsets up to the PDH locking bandwidth. Notably, the high internal quality factor of the ECTL, due to the low-loss ring resonators, provides high resilience to optical feedback and removes the need for an optical isolator between the ECTL and coil reference cavity to operate the stabilization lock. Eliminating the need for an isolator between the two components is important for future integration of the laser and reference cavity onto a single silicon nitride chip.

The stabilized-ECTL frequency noise, linewidth, and Allan Deviation (ADEV) are measured using two independent techniques. For FN above 3 kHz frequency offset we use an asymmetric fiber Mach-Zehnder interferometer (MZI) optical frequency discriminator (OFD). Below 3 kHz offset down to 1 Hz we measure the noise of a heterodyne beatnote of the ECTL with an optical frequency comb that has been locked to a stable reference laser (SRL). The beatnote is then measured on a precision frequency counter (See Methods for more details). The FN of the coil-locked ECTL operating at 1550 nm is plotted in Fig. 3a. The stabilized ECTL (orange) has a FN reduction of more than 6-orders of magnitude at low frequency offsets compared with the free-running laser (blue) and reaches the TRN limit of the 10-m coil resonator (red, dashed) between 1-100 kHz offset, before sloping upwards due to the servo bump corresponding to the PDH locking bandwidth at around 0.5 MHz. The measured 1/π integral linewidth of the stabilized-ECTL at 1550 nm is 27 Hz, reduced from 1750 Hz for the free-running laser, an integral linewidth reduction of 65x. The minimum FN is 0.12 $Hz^2$/Hz at 16 kHz, and the corresponding fractional frequency stability, or Allan deviation (Fig. 3b), has a minimum of $1.8 \times 10^{-13}$ at 6.4 ms and a drift of 5 kHz/sec. The fundamental and integral linewidths of the stabilized-ECTL at various points across the tuning range are plotted in Fig. 3c, see Supplementary Note 4 for more details.

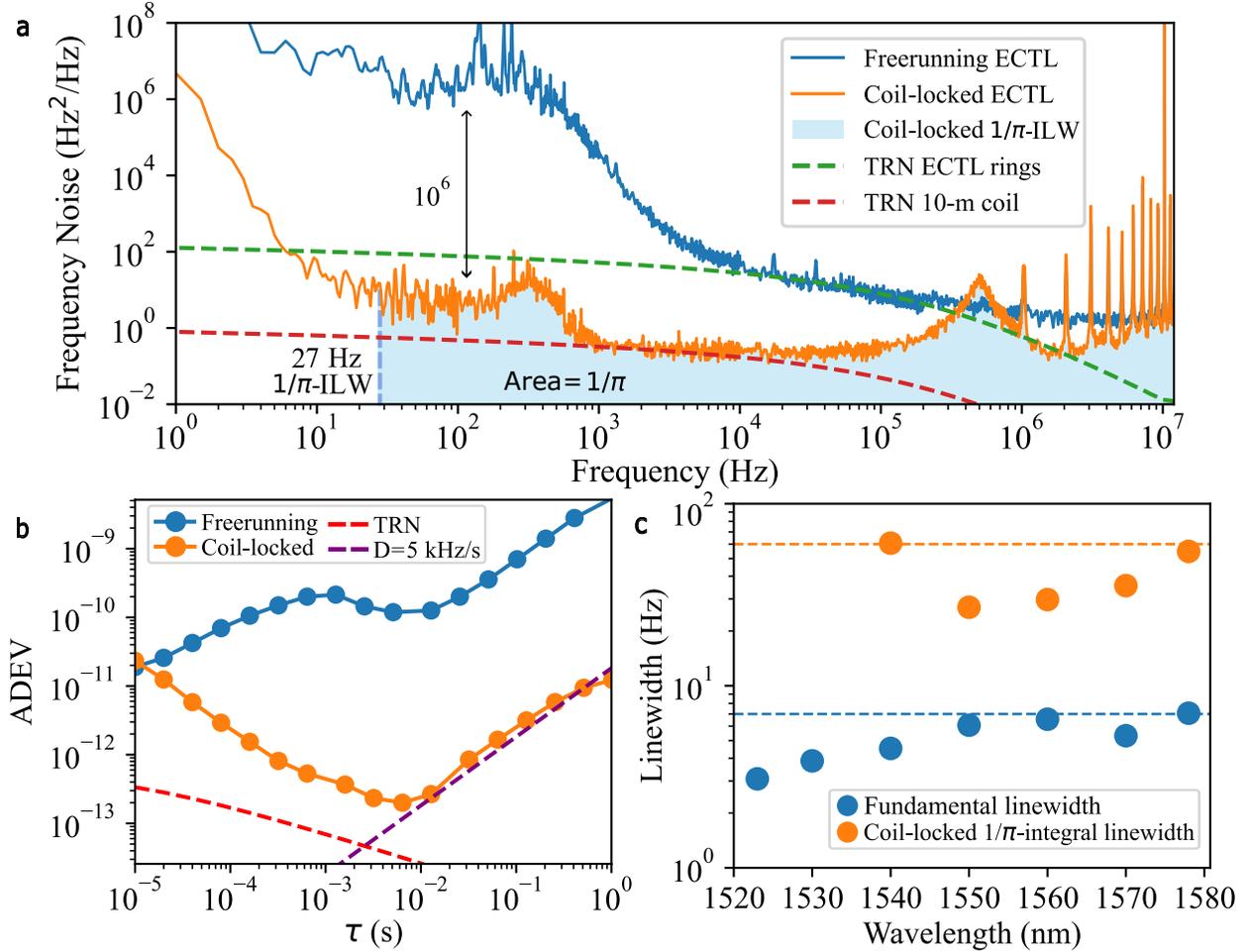

**Fig. 3 Coil-stabilized ECTL frequency noise, stability, and fundamental and integral linewidths across tuning range. a** An example frequency noise (FN) spectrum of the free-running (blue) and coil-locked (orange) external cavity tunable laser (ECTL) at 1550 nm shows > 6 orders of magnitude reduction in FN at low frequency offsets. The shaded light-blue area corresponds to the $1/\pi$ reverse integral linewidth (ILW) of the locked laser measured at 27 Hz. The minimum FN of the locked laser is measured at 0.12 Hz$^2$/Hz at 16 kHz frequency offset. The Pound-Drever-Hall (PDH) lock servo bump is indicated at 0.5 MHz. **b** Allan deviation (ADEV) of the free-running (blue) and coil-locked (orange) ECTL demonstrating 1.8x10$^{-13}$ at 6.4 ms and 5 kHz/s drift. **c** The measured fundamental linewidth (FLW) (blue) across the 60 nm tuning range and coil-locked ILW (orange) across a 40 nm tuning range. The ILW measurement was limited by the beatnote measurement below 1540 nm. In addition, the coil-lock at 1520 nm and below was limited due to low extinction ratio (ER). The dashed lines indicate the highest measured value of 7 Hz for FLW (blue) and 60 Hz for ILW (orange).

**ECTL resilience to optical feedback.** The ECTL resilience to optical feedback is demonstrated by operating the PDH lock to the 10-meter coil reference cavity without an isolator – marking the first such demonstration for a Vernier-style laser and a key step toward fully integrated stabilized lasers on-chip. To further investigate this under controlled conditions, we utilized an optical feedback measurement setup[16,43-45] (see Fig. 4a). Using a fiber circulator and a variable optical attenuator (VOA) we precisely control the feedback level back to the laser while accounting for circuit losses to determine the power reaching the laser. A small fraction of the ECTL output is diverted to measure the power and the frequency noise of the laser, the results of which are plotted in Fig. 4b. Compared to a commercial III-V DFB laser that experiences FN degradation and coherence collapse at a feedback level of -40 dB[16,43-45], the ECTL maintains single-mode operation across all tested feedback levels, up to -10 dB, with no degradation in the frequency noise. This represents a 30 dB improvement in resilience to optical feedback over a commercial DFB laser and enables the isolator-free coil resonator stabilization.

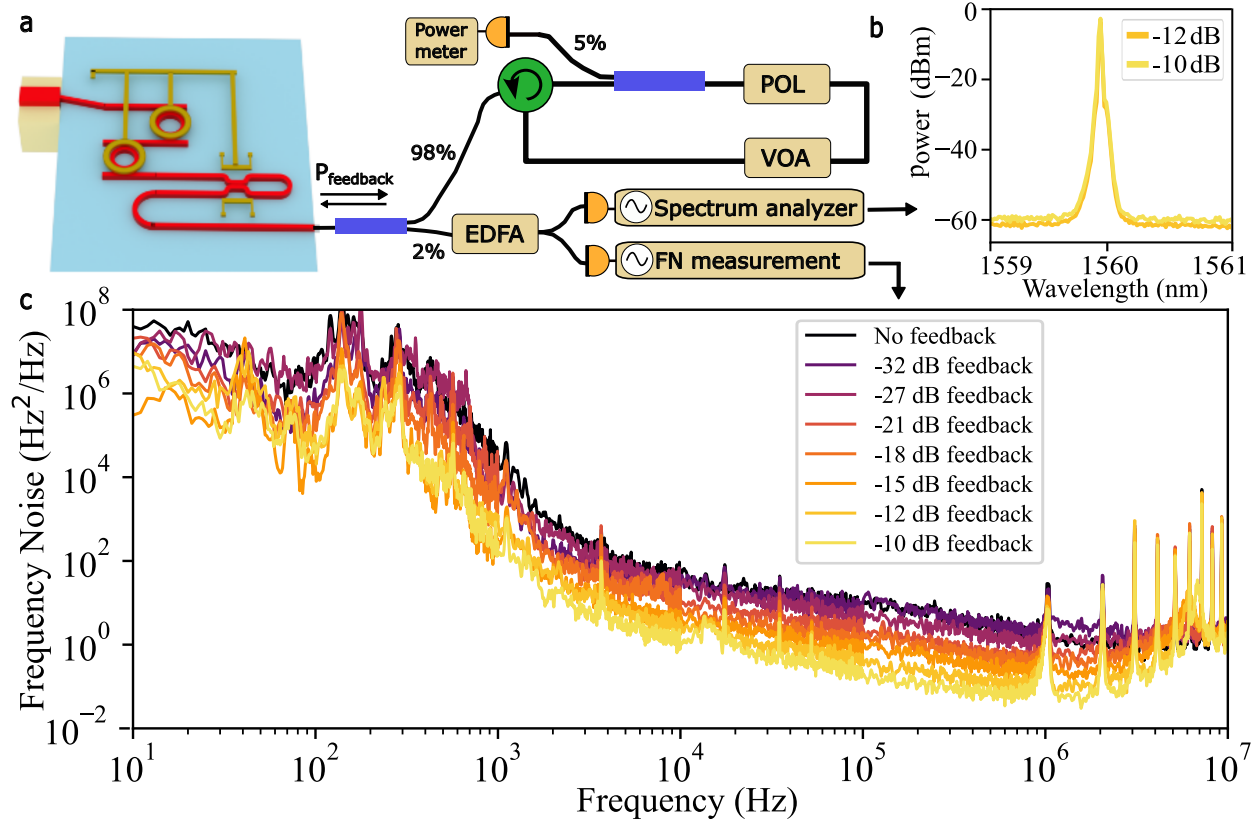

**Fig. 4 ECTL feedback measurements. a** Schematic of the experimental setup for measuring the effect of optical feedback on the external cavity tunable laser (ECTL). The laser output power is initially split in two: the majority for feeding back to the laser and a small portion to measure the laser frequency noise (FN). A variable optical attenuator (VOA) sets the feedback power level that is monitored with a power meter. Polarization (POL) paddles ensure the returned optical field is aligned properly with the silicon nitride waveguide. The FN is measured with an optical frequency discriminator (OFD) and an optical spectrum analyzer (OSA) monitors for the onset of coherence collapse. **b** OSA trace of the ECTL output under the two highest feedback levels of -12 and -10 dB. **c** Frequency noise plots of the free running ECTL operating at 1560 nm under different optical feedback levels ranging from -32 to -10 dB. The maximum feedback level is limited by the fiber-chip insertion loss and all fiber splitter and connector losses. The inset plots the OSA traces for ECTL output under the two highest optical feedback levels showing single mode operation.

We observe in Fig. 4b that the measured frequency noise of the ECTL decreases with increasing optical feedback. This is likely due to the long coherence time and strong mode selectivity of the ECTL where photons that re-enter from the external feedback loop remain phase-coherent with the intracavity field, effectively increasing the intracavity photon number and reducing quantum noise. Additionally, the extended fiber feedback circuit increases the overall optical mode volume, which suppresses TRN, leading to a lower frequency noise floor.

# DISCUSSION

We demonstrate a precision ultra-low instantaneous and integral linewidth widely tunable stabilized hybrid integrated ECTL locked to an integrated coil reference cavity that achieves record frequency and phase noise for hybrid-integrated widely tunable lasers. The coil-locked ECTL realizes 6 Hz fundamental linewidth with 27 Hz integral linewidth and $1.8 \times 10^{-13}$ fractional frequency stability at 1550 nm, and 3 - 7 Hz fundamental with sub-60 Hz integral linewidths measured across the tuning range. These linewidths are the lowest to date over the widest tuning range for integrated chips. Furthermore, we demonstrate isolation free stabilization of the ECTL to the coil cavity, enabled by the high Q of the ECTL cavity. The 10-meter coil has an FSR of 20 MHz which allows locking and stabilization at almost all wavelengths across the tuning range, a distinct advantage over low FSR bulk optic reference cavities.

We compare our results to the state of the art in Table 1 and Fig. 5 with other hybrid-integrated low noise lasers. The table is sparsely populated in several columns since this work represents one of the few that report the $1/\pi$ integral linewidth which is important for many precision applications. When not available directly in the publication we have calculated the ILW from the available FN data in that publication. We believe this comparison highlights the unique properties of the hybrid-integrated ECTL locked to the integrated coil reference cavity. For example, we report both the fundamental and integral linewidths over a full tuning range. The fundamental linewidth is 1.5X lower and the $1/\pi$ integral linewidth 3 orders of magnitude lower with a tuning range 1.5X larger than state of the art integrated ECTLs[26]. Additionally, this work is the first to report an ADEV, 1.6E-13 at 6.4 ms, and isolator free operation in a Vernier-style hybrid laser.

**Table 1. Comparison of low linewidth hybrid-integrated tunable lasers**

| Platform | Laser type | λ (nm) | FLW (Hz) | ILW (Hz) 1/π | ADEV (@ 1 ms) | Tuning (nm) | Output power (mW) | SMSR (dB) | Optical isolation (dB) |
|---|---|---|---|---|---|---|---|---|---|
| $Si_3N_4$ [16] | SIL | 1550 | 0.04 | 236* | … | 0.8 | 0.3 | 60 | … |
| $Si_3N_4$ [49] | SIL | 1550 | 3.8 | 4,715* | … | … | 10.5 | 65 | … |
| $Si_3N_4$ [50] | SIL | 1550 | 3 | 1,560* | … | … | … | 54 | … |
| $Si_3N_4$ [18] | SIL | 780 | 0.74 | 864 | … | 2 | 2 | 36 | … |
| $Si_3N_4$ [51] | SIL | 785 | 700 | 50,173* | … | 12 | 10 | 37 | … |
| $Si_3N_4$ [23] | EDBR | 1550 | 320 | 47,466* | … | … | 24 | 55 | … |
| Si [52] | ECTL: 3 ring | 1550 | 220 | 33,246* | … | 110 | 3 | 50 | … |
| Si [28] | ECTL: 3 ring | 1550 | 95 | 9,237* | … | 120 | 1.5 | 60 | … |
| $Si_3N_4$ [39] | ECTL: 3 ring | 1550 | 40 | 87,844** | … | 70 | 23 | 60 | … |
| $Si_3N_4$ [53] | ECTL: 2 ring | 1550 | 2,200 | 57,526* | … | 120 | 24 | 63 | … |
| $Si_3N_4$ [36] | ECTL: 2 ring | 1550 | 750 - 4,000 | 31,614* | … | 172 | 26 | 68 | … |
| $Si_3N_4$ [54] | ECTL: 2 ring | 852 | 65 | 6,770* | … | 15 | 25 | 50 | … |
| $Si_3N_4$ [26] | ECTL: 2 ring | 1550 | 6 - 9.8*** | 2,350 | … | 40 | 4.8 | 64 | … |
| Si$_3$N$_4$ [†] | ECTL (2 ring) + coil | 1550 | 3 - 7[***] | 27 | 1.8E-13 @ 6.4 ms | 60 | 4.4 | 65 | -30[****] |

[†] This work
[*] Not reported in manuscript: calculated from published FN data
[**] ILW limited by available FN data
[***] Measured across tuning range
[****] Relative to a typical commercial III-V DFB laser

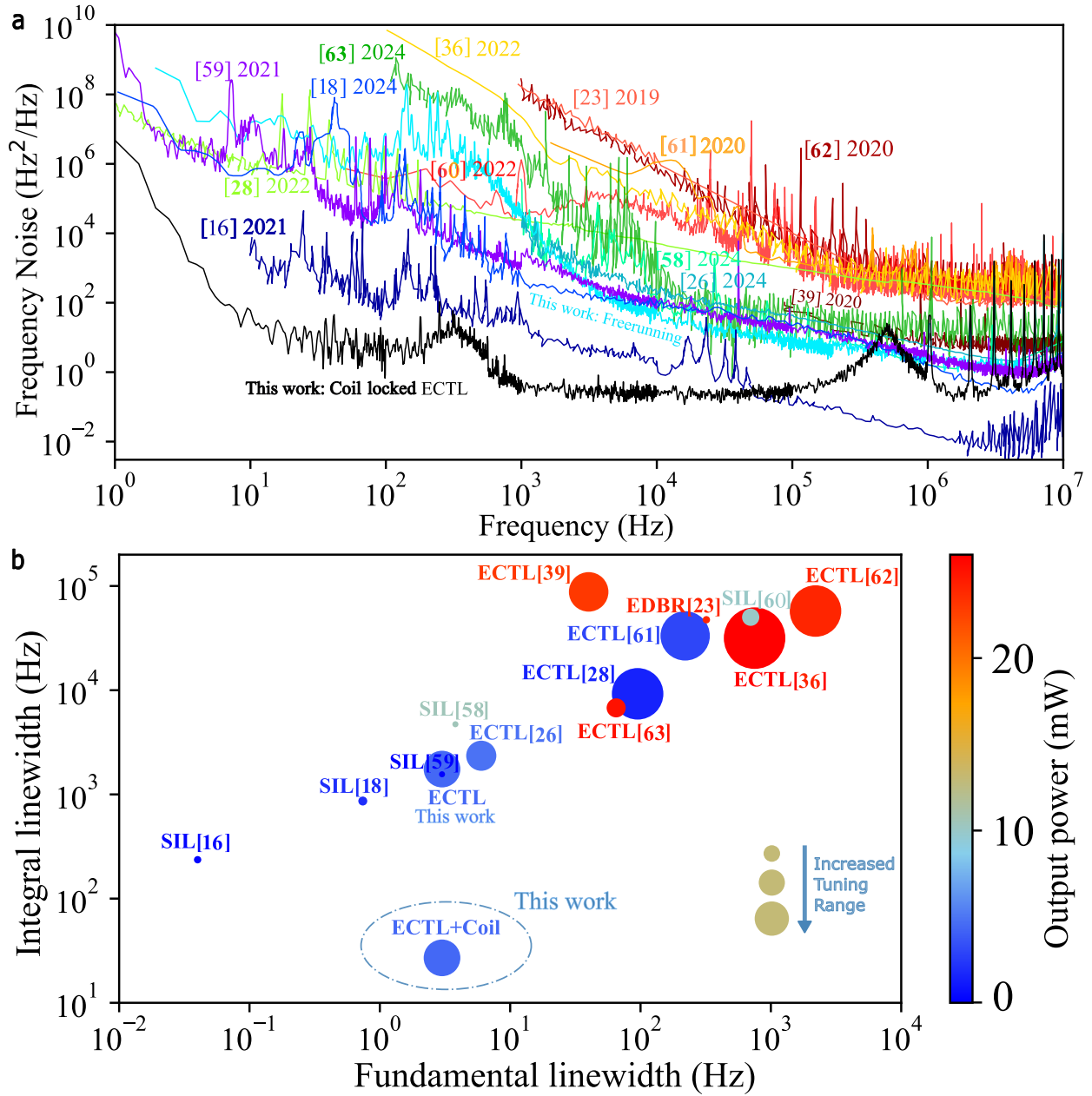

**Fig. 5 Comparison of low linewidth hybrid-integrated lasers summarized in Table 1. a** Frequency noise plots comparing hybrid-integrated low noise lasers. **b** Fundamental vs. 1/π integral linewidths. The bubble size represents the tuning range and color heat map represents output power.

Since both the ECTL and coil resonator are fabricated in the same 80 nm thick, CMOS-compatible $Si_3N_4$ platform, these results present a clear path forward towards realizing a fully integrated, chip-scale stabilized laser that combines narrow instantaneous linewidth with low frequency drift in one device. In future versions, the EOM used in this experiment can be eliminated by adding a PZT-on-$Si_3N_4$ integrated double sideband modulator that can achieve locking bandwidths up to 20 MHz[55], or by employing modulation free stabilization techniques[55,56]. Thermal and PZT actuators have been demonstrated in this $Si_3N_4$ platform without affecting waveguide loss, independent of wavelength, and operate from DC out to 10s of MHz[35] and could also be used to directly tune the intracavity ECTL rings rather than feeding the PDH error signal back to the gain chip current. Additionally, the core components of this stabilized laser design have been demonstrated across the visible to SWIR range[57] making this hybrid-integrated stabilized laser design a potential laser source for silicon nitride photonics that support atomic and quantum applications[58-60].

Further experiments and design improvements will yield better laser performance in terms of tunability, variation in integral and fundamental linewidth, and higher output power. For example, full control of the Sagnac loop mirror would provide an adjustment of the laser output power, and therefore also the intracavity power, allowing the user to optimize for higher output power versus lower fundamental linewidth. Additionally, loaded-Qs as high as 100 million have been demonstrated in this platform, and since the modified Schawlow-Townes laser linewidth will decrease as $1/Q^2$, we predict that future devices could achieve even lower fundamental linewidths. Future designs can also include coil resonators longer than 10-meters to increase the mode volume and further reduce the TRN and integral linewidth, as well as incorporating tunable coupling[61] to optimize the laser locking conditions across a wide range of wavelengths using a single resonator. For applications that require high optical output power additional gain blocks can be added and operated in parallel with a shared high-Q silicon nitride external cavity[25] where output powers >100 mW have already been demonstrated in a dual-gain hybrid-integrated laser[46]. Other pathways to increasing the output power are to incorporate on-chip amplifiers[62,63] or through injection-locked amplification.

## METHODS

**Fabrication process.** The lower cladding consists of a 15-μm-thick thermal oxide grown on a 100 mm diameter, 1 mm thick silicon wafer substrate. The main waveguide layer is an 80 nm thick stoichiometric $Si_3N_4$ film deposited on the lower cladding using low-pressure chemical vapor deposition (LPCVD). A PAS 5500 ASML deep ultraviolet (DUV) stepper was used to pattern a DUV photoresist layer. The high-aspect-ratio waveguide core is formed by anisotropically dry etching the $Si_3N_4$ film in a Panasonic E626I Inductively Coupled Plasma-Reactive Ion etcher (ICP-RIE) using a $CHF_3/CF_4/O_2$ chemistry. After the etch, the wafer is cleaned using a standard RCA cleaning process. A 5-μm-thick silicon dioxide upper cladding layer was deposited using plasma-enhanced chemical vapor deposition (PECVD) with tetraethoxysilane (TEOS) as a precursor. This is followed by a final two-step anneal at 1050 °C for 7 hours and 1150 °C for 2 hours which is an optimized anneal process for our waveguides.

**RSOA coupling.** The gain chip is coated with 90% reflectivity on the side opposite the silicon nitride PIC and an antireflection (AR) material on the near side with a reflectivity of 0.005%. It is then mounted on a temperature-controlled copper block for heat-sinking. The gain chip is wire bonded to a PCB that screws onto the copper block for external electrical control of the gain chip. The SOA has an angled facet of 5.6 degrees that requires, based on Snell's Law, the $Si_3N_4$ waveguide to be angled by 13.1 degrees to best match the beam propagation direction. The $Si_3N_4$ waveguide is initially 18 μm-wide to achieve an estimated optimal mode overlap of 52%. We estimate that the actual coupling loss is around 4-5 dB. The $Si_3N_4$ PIC sits atop its own temperature-controlled mount and the RSOA is edge-coupled to the ECTL input waveguide.

**Frequency noise measurements.** The coil-stabilized ECTL frequency noise, linewidth, and stability is measured using two independent techniques[20,21,30,68,69]. For FN above 3 kHz frequency offset we use an unbalanced fiber-MZI with a 1.026 MHz FSR as an optical frequency discriminator (OFD) and measure the self-delayed homodyne FN signal on a high-speed balanced photodetector. Below 3 kHz offset the OFD fiber-MZI noise can become dominant so we instead photomix the ECTL with a stable reference laser (SRL) and measure the heterodyne beatnote signal with a Keysight 53230A precision frequency counter. The SRL system consists of a single frequency Rock fiber laser PDH-locked to a Stable Laser Systems (SLS) 1550 nm ultralow expansion (ULE) cavity that delivers Hz-level linewidth and ~0.1 Hz/s frequency drift. Additionally, we use a Vescent self-referenced fiber frequency comb with a 100 MHz $f_{rep}$ and lock it to the SLS to extend the stability of the SRL system to many wavelengths and enable accurate close-to-carrier FN measurements across the ECTL spectrum.

**Optical feedback measurements.** The optical feedback measurement, illustrated in Fig. 4a, utilizes a fiber circulator and a variable optical attenuator (VOA) to control the optical feedback level. The power returning to the PIC is polarization sensitive and must be adjusted to maximize the optical feedback before making any measurements. The power returned to the ECTL is monitored by tapping 5% of the feedback power and measuring on a power meter. Another small fraction of the laser power is diverted to an EDFA and then to an OFD setup to measure the frequency noise of the laser under different feedback conditions. The full fiber feedback circuit is ~25 meters long. All losses in the feedback circuit must be accounted for to most accurately determine the actual feedback power to the laser. The highest level of optical feedback in our experiment was -10 dB, limited primarily by the optical losses due to fiber-to-chip coupling and the connectors and splitters in the feedback loop.

# DATA AVAILABILITY

The data that support the plots within this paper and other findings of this study are available from the corresponding author on request.

# REFERENCES

1. Ludlow, A. D., Boyd, M. M., Ye, J., Peik, E. & Schmidt, P. O. Optical atomic clocks. *Reviews of Modern Physics* **87**, 637-701 (2015). https://doi.org/10.1103/RevModPhys.87.637
2. Norcia, M. A. *et al.* Seconds-scale coherence on an optical clock transition in a tweezer array. *Science* **366**, 93 (2019). https://doi.org/10.1126/science.aay0644
3. Ladd, T. D. *et al.* Quantum computers. *Nature* **464**, 45-53 (2010). https://doi.org/10.1038/nature08812
4. Bluvstein, D. *et al.* Logical quantum processor based on reconfigurable atom arrays. *Nature* **626**, 58-65 (2024). https://doi.org/10.1038/s41586-023-06927-3
5. Graham, T. M. *et al.* Multi-qubit entanglement and algorithms on a neutral-atom quantum computer. *Nature* **604**, 457-462 (2022). https://doi.org/10.1038/s41586-022-04603-6
6. Pino, J. M. *et al.* Demonstration of the trapped-ion quantum CCD computer architecture. *Nature* **592**, 209-213 (2021). https://doi.org/10.1038/s41586-021-03318-4
7. Abramovici, A. *et al.* LIGO: The Laser Interferometer Gravitational-Wave Observatory. *Science* **256**, 325-333 (1992). https://doi.org/10.1126/science.256.5055.325
8. Ye, J., Kimble, H. J. & Katori, H. Quantum State Engineering and Precision Metrology Using State-Insensitive Light Traps. *Science* **320**, 1734-1738 (2008). https://doi.org/10.1126/science.1148259
9. Degen, C. L., Reinhard, F. & Cappellaro, P. Quantum sensing. *Reviews of Modern Physics* **89**, 035002-035001 - 035002-035039 (2017). https://doi.org/10.1103/RevModPhys.89.035002
10. Fan, H. *et al.* Atom based RF electric field sensing. *Journal of Physics B: Atomic, Molecular and Optical Physics* **48**, 202001 (2015). https://doi.org/10.1088/0953-4075/48/20/202001
11. Marra, G. *et al.* Ultrastable laser interferometry for earthquake detection with terrestrial and submarine cables. *Science* **361**, 486-490 (2018). https://doi.org/10.1126/science.aat4458
12. Sun, S. *et al.* Integrated optical frequency division for microwave and mmWave generation. *Nature* (2024). https://doi.org/10.1038/s41586-024-07057-0
13. Kudelin, I. *et al.* Photonic chip-based low-noise microwave oscillator. *Nature* (2024). https://doi.org/10.1038/s41586-024-07058-z
14. Matei, D. G. *et al.* 1.5 um Lasers with Sub-10 mHz Linewidth. *Physical Review Letters* **118**, 263202 (2017). https://doi.org/10.1103/PhysRevLett.118.263202
15. Hirata, S., Akatsuka, T., Ohtake, Y. & Morinaga, A. Sub-hertz-linewidth diode laser stabilized to an ultralow-drift high-finesse optical cavity. *Applied Physics Express* **7**, 022705 (2014). https://doi.org/10.7567/apex.7.022705
16. Li, B. *et al.* Reaching fiber-laser coherence in integrated photonics. *Optics Letters* **46**, 5201-5204 (2021). https://doi.org/10.1364/OL.439720
17. Jin, W. *et al.* Hertz-linewidth semiconductor lasers using CMOS-ready ultra-high-Q microresonators. *Nature Photonics* **15**, 346-353 (2021). https://doi.org/10.1038/s41566-021-00761-7
18. Isichenko, A. *et al.* Sub-Hz fundamental, sub-kHz integral linewidth self-injection locked 780 nm hybrid integrated laser. *Scientific Reports* **14**, 27015 (2024). https://doi.org/10.1038/s41598-024-76699-x
19. Li, B. *et al.* High-coherence hybrid-integrated 780 nm source by self-injection-locked second-harmonic generation in a high-Q silicon-nitride resonator. *Optica* **10**, 1241-1244 (2023). https://doi.org/10.1364/OPTICA.498391
20. Gundavarapu, S. *et al.* Sub-hertz fundamental linewidth photonic integrated Brillouin laser. *Nature Photonics* **13**, 60-67 (2019). https://doi.org/10.1038/s41566-018-0313-2

21    Li, J., Lee, H., Chen, T. & Vahala, K. J. Characterization of a high coherence, Brillouin microcavity laser on silicon. *Opt. Express* **20**, 20170-20180 (2012). https://doi.org/10.1364/OE.20.020170
22    Chauhan, N. *et al.* Visible light photonic integrated Brillouin laser. *Nature Communications* **12**, 4685 (2021). https://doi.org/10.1038/s41467-021-24926-8
23    Xiang, C., Morton, P. A. & Bowers, J. E. Ultra-narrow linewidth laser based on a semiconductor gain chip and extended Si3N4 Bragg grating. *Optics Letters* **44**, 3825-3828 (2019). https://doi.org/10.1364/OL.44.003825
24    Santis, C. T., Steger, S. T., Vilenchik, Y., Vasilyev, A. & Yariv, A. High-coherence semiconductor lasers based on integral high-Q resonators in hybrid Si/III-V platforms. *Proceedings of the National Academy of Sciences* **111**, 2879-2884 (2014). https://doi.org/10.1073/pnas.1400184111
25    Zhao, R. *et al.* Hybrid dual-gain tunable integrated InP-Si3N4 external cavity laser. *Opt. Express* **29**, 10958-10966 (2021). https://doi.org/10.1364/OE.416398
26    Wu, Y. *et al.* Hybrid integrated tunable external cavity laser with sub-10 Hz intrinsic linewidth. *APL Photonics* **9**, 021302 (2024). https://doi.org/10.1063/5.0190696
27    Ren, Y., Xiong, B., Yu, Y., Lou, K. & Chu, T. Widely and fast tunable external cavity laser on the thin film lithium niobate platform. *Optics Communications* **559**, 130415 (2024). https://doi.org/https://doi.org/10.1016/j.optcom.2024.130415
28    Morton, P. A. *et al.* Integrated Coherent Tunable Laser (ICTL) With Ultra-Wideband Wavelength Tuning and Sub-100 Hz Lorentzian Linewidth. *Journal of Lightwave Technology* **40**, 1802-1809 (2022). https://doi.org/10.1109/JLT.2021.3127155
29    Blumenthal, D. J., Heideman, R., Geuzebroek, D., Leinse, A. & Roeloffzen, C. Silicon Nitride in Silicon Photonics. *Proceedings of the IEEE* **106**, 2209-2231 (2018). https://doi.org/10.1109/JPROC.2018.2861576
30    Puckett, M. W. *et al.* 422 Million intrinsic quality factor planar integrated all-waveguide resonator with sub-MHz linewidth. *Nature Communications* **12**, 934 (2021). https://doi.org/10.1038/s41467-021-21205-4
31    Liu, K. *et al.* Ultralow 0.034 dB/m loss wafer-scale integrated photonics realizing 720 million Q and 380 μW threshold Brillouin lasing. *Optics Letters* **47**, 1855-1858 (2022). https://doi.org/10.1364/OL.454392
32    Park, H., Zhang, C., Tran, M. A. & Komljenovic, T. Heterogeneous silicon nitride photonics. *Optica* **7**, 336-337 (2020). https://doi.org/10.1364/OPTICA.391809
33    Tian, H. *et al.* Hybrid integrated photonics using bulk acoustic resonators. *Nature Communications* **11**, 3073 (2020). https://doi.org/10.1038/s41467-020-16812-6
34    Dong, M. *et al.* High-speed programmable photonic circuits in a cryogenically compatible, visible–near-infrared 200 mm CMOS architecture. *Nature Photonics* **16**, 59-65 (2022). https://doi.org/10.1038/s41566-021-00903-x
35    Wang, J., Liu, K., Harrington, M. W., Rudy, R. Q. & Blumenthal, D. J. Silicon nitride stress-optic microresonator modulator for optical control applications. *Opt. Express* **30**, 31816-31827 (2022). https://doi.org/10.1364/OE.467721
36    Guo, Y. *et al.* Hybrid integrated external cavity laser with a 172-nm tuning range. *APL Photonics* **7**, 066101 (2022). https://doi.org/10.1063/5.0088119
37    Fan, Y. *et al.* in *Conference on Lasers and Electro-Optics.*  JTh5C.9 (Optical Society of America).
38    Hao, L. *et al.* Narrow-linewidth self-injection locked diode laser with a high-Q fiber Fabry–Perot resonator. *Optics Letters* **46**, 1397-1400 (2021). https://doi.org/10.1364/OL.415859
39    Fan, Y. *et al.* Hybrid integrated InP-Si3N4 diode laser with a 40-Hz intrinsic linewidth. *Opt. Express* **28**, 21713-21728 (2020). https://doi.org/10.1364/OE.398906
40    Liu, K. *et al.* 36 Hz integral linewidth laser based on a photonic integrated 4.0 m coil resonator. *Optica* **9**, 770-775 (2022). https://doi.org/10.1364/OPTICA.451635
41    Lee, H. *et al.* Spiral resonators for on-chip laser frequency stabilization. *Nature Communications* **4**, 2468 (2013). https://doi.org/10.1038/ncomms3468

42  Huang, G. *et al.* Thermorefractive noise in silicon-nitride microresonators. *Physical Review A* **99**, 061801 (2019). https://doi.org/10.1103/PhysRevA.99.061801
43  Harfouche, M. *et al.* Kicking the habit/semiconductor lasers without isolators. *Opt. Express* **28**, 36466-36475 (2020). https://doi.org/10.1364/OE.411816
44  Zhang, Z. *et al.* High-Speed Coherent Optical Communication With Isolator-Free Heterogeneous Si/III-V Lasers. *Journal of Lightwave Technology* **38**, 6584-6590 (2020). https://doi.org/10.1109/JLT.2020.3015738
45  Xiang, C. *et al.* 3D integration enables ultralow-noise isolator-free lasers in silicon photonics. *Nature* **620**, 78-85 (2023). https://doi.org/10.1038/s41586-023-06251-w
46  White, A. D. *et al.* Unified laser stabilization and isolation on a silicon chip. *Nature Photonics* **18**, 1305-1311 (2024). https://doi.org/10.1038/s41566-024-01539-3
47  Tang, L., Li, J., Yang, S., Chen, H. & Chen, M. A Method for Improving Reflection Tolerance of Laser Source in Hybrid Photonic Packaged Micro-System. *IEEE Photonics Technology Letters* **33**, 465-468 (2021). https://doi.org/10.1109/LPT.2021.3069220
48  Kondratiev, N. M. & Gorodetsky, M. L. Thermorefractive noise in whispering gallery mode microresonators: Analytical results and numerical simulation. *Physics Letters A* **382**, 2265-2268 (2018). https://doi.org/https://doi.org/10.1016/j.physleta.2017.04.043
49  Siddharth, A. *et al.* Piezoelectrically tunable, narrow linewidth photonic integrated extended-DBR lasers. *Optica* **11**, 1062-1069 (2024). https://doi.org/10.1364/OPTICA.524703
50  Xiang, C. *et al.* High-performance lasers for fully integrated silicon nitride photonics. *Nature Communications* **12**, 6650 (2021). https://doi.org/10.1038/s41467-021-26804-9
51  Corato-Zanarella, M. *et al.* Widely tunable and narrow-linewidth chip-scale lasers from near-ultraviolet to near-infrared wavelengths. *Nature Photonics* (2022). https://doi.org/10.1038/s41566-022-01120-w
52  Tran, M. A. *et al.* Ring-Resonator Based Widely-Tunable Narrow-Linewidth Si/InP Integrated Lasers. *IEEE Journal of Selected Topics in Quantum Electronics* **26**, 1-14 (2020). https://doi.org/10.1109/JSTQE.2019.2935274
53  van Rees, A. *et al.* Ring resonator enhanced mode-hop-free wavelength tuning of an integrated extended-cavity laser. *Opt. Express* **28**, 5669-5683 (2020). https://doi.org/10.1364/OE.386356
54  Nejadriahi, H. *et al.* Sub-100 Hz intrinsic linewidth 852 nm silicon nitride external cavity laser. *Optics Letters* **49**, 7254-7257 (2024). https://doi.org/10.1364/OL.543307
55  Wang, J., Liu, K., Rudy, R. Q. & Blumenthal, D. J. in *Optica Quantum 2.0 Conference and Exhibition.* QW3B.3 (Optica Publishing Group).
56  Idjadi, M. H., Kim, K. & Fontaine, N. K. Modulation-free laser stabilization technique using integrated cavity-coupled Mach-Zehnder interferometer. *Nature Communications* **15**, 1922 (2024). https://doi.org/10.1038/s41467-024-46319-3
57  Chauhan, N. *et al.* Ultra-low loss visible light waveguides for integrated atomic, molecular, and quantum photonics. *Opt. Express* **30**, 6960-6969 (2022). https://doi.org/10.1364/OE.448938
58  Isichenko, A. *et al.* Photonic integrated beam delivery for a rubidium 3D magneto-optical trap. *Nature Communications* **14**, 3080 (2023). https://doi.org/10.1038/s41467-023-38818-6
59  Niffenegger, R. J. *et al.* Integrated multi-wavelength control of an ion qubit. *Nature* **586**, 538-542 (2020). https://doi.org/10.1038/s41586-020-2811-x
60  Mehta, K. K. *et al.* Integrated optical multi-ion quantum logic. *Nature* **586**, 533-537 (2020). https://doi.org/10.1038/s41586-020-2823-6
61  Liu, K. *et al.* Tunable broadband two-point-coupled ultra-high-Q visible and near-infrared photonic integrated resonators. *Photon. Res.* **12**, 1890-1898 (2024). https://doi.org/10.1364/PRJ.528398
62  Op de Beeck, C., Elsinger, L., Haq, B., Roelkens, G. & Kuyken, B. in *Frontiers in Optics + Laser Science APS/DLS.* FTu6B.1 (Optica Publishing Group).
63  Mu, J., Dijkstra, M., Korterik, J., Offerhaus, H. & García-Blanco, S. M. High-gain waveguide amplifiers in Si3N4 technology via double-layer monolithic integration. *Photon. Res.* **8**, 1634-1641 (2020). https://doi.org/10.1364/PRJ.401055

## ACKNOWLEDGEMENTS

We acknowledge Henry Timmers at Vescent for help with the setup of the fiber frequency comb and Mark W. Harrington at UCSB for help with the SRL system and Karl Nelson of Honeywell for help fabricating the coil resonator. This work is based in part by funding from DARPA GRYPHON award HR0011-22-2-0008. The views and conclusions contained in this document are those of the author(s) and should not be interpreted as representing the official policies of DARPA or the U.S. government.

## AUTHOR CONTRIBUTIONS

DASH, DJB, KL, DB and AI prepared the manuscript. DB designed and fabricated the ECTL PICs. DASH and AI built the ECTL. DASH characterized the ECTL performance. KL designed and tested the coil resonator. DASH and KL performed all laser locking and frequency noise experiments. AI built the SRL and optical frequency comb frequency noise measurement system. DJB supervised the project. All authors participated in writing of the manuscript.

## COMPETING INTERESTS

Dr. Blumenthal's work has been funded by ColdQuanta d.b.a. Infleqtion. Dr. Blumenthal has consulted for Infleqtion and received compensation, is a member of the scientific advisory council, and owns stock in the company. D. A. S. Heim, D. Bose, K. Liu, and A. Isichenko declare no competing interests.

# Supplemental information

## Supplementary Note 1: Introduction

This supplemental section provides additional information on the design and performance of the hybrid-integrated external cavity tunable laser (ECTL), including a diagram of the working principle of the dual-ring ECTL, measurement of the ring resonator Qs, measurements of the 10-meter coil resonator, more information regarding the free-running and integrated coil-stabilized frequency noise measurements, as well as a calculation of the optical feedback parameter.

The basic working principle of the dual-ring ECTL is illustrated in Supplementary Fig. 1a. The RSOA has a broad power spectrum and the two ring resonators, with slightly different radii, provide linewidth narrowing where the two resonances overlap. Changing the resonance of one of the rings adjusts which modes overlap and thereby tunes the laser up to one Vernier FSR. The back mirror of the RSOA and the loop mirror of the external $Si_3N_4$ circuit form the extended laser cavity and has associated Fabry Perot modes.

## Supplementary Note 2: ECTL design

The layer stack for the low-loss $Si_3N_4$ waveguide platform is indicated in Supplementary Fig. 1b. The ECTL waveguide dimension is 2.6 µm by 80 nm resulting in a dilute optical mode to reduce scattering losses. Measurement of the quality factor (Q) of the intracavity rings, shown in Supplementary Fig. 1c, yield loaded and intrinsic-Qs of 0.6 and 3.5 million, respectively. The rings were intentionally heavily coupled to reduce cavity coupling losses to lower the lasing threshold at the expense of having higher loaded-Qs and longer photon lifetimes. This is a design parameter that can be optimized in future designs.

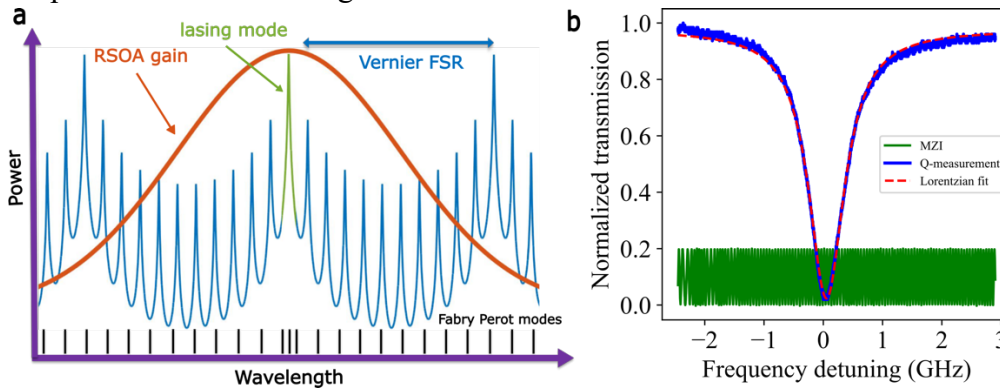

**Supplementary Fig. 1 ECTL working principle and characterization. a** An illustration of the basic working principle of the dual-ring external cavity tunable laser (ECTL). **b** Q-measurement of one of the ECTL rings (blue) fitted to a Lorentzian function (red) to extract Q values of 0.65 million loaded and 3.5 million intrinsic. An 18 MHz fiber-MZI (green) is used to calibrate the frequency detuning.

## Supplementary Note 3: 10-meter-coil resonator design and characterization

The 10-meter coil resonator has a waveguide dimension of 6 µm by 80 nm utilizing the fundamental TE mode. The coil spiral center-to-center waveguide spacing is 25 µm, and the circular S-bend diameter in the center of the coil waveguides is ~3.6 mm, resulting in a much smaller device size and making it possible to wrap the 10-meter-coil waveguide on a single die size (21.6 mm by 26 mm). The bus-resonator directional coupler uses a 2.5 µm coupler gap and 1.5 mm coupling length. The bus waveguide is tapered from 6 µm to 1.5 µm for better fiber-to-chip edge coupling. The fiber-pigtailed device is packaged with a metal enclosure for better handling capabilities, shown in Supplementary Fig. 2a. The FSR is measured to be around 19.4 MHz, and the intrinsic Q reaches above 300 million around 1600 nm, shown in Supplementary Fig. 2d.

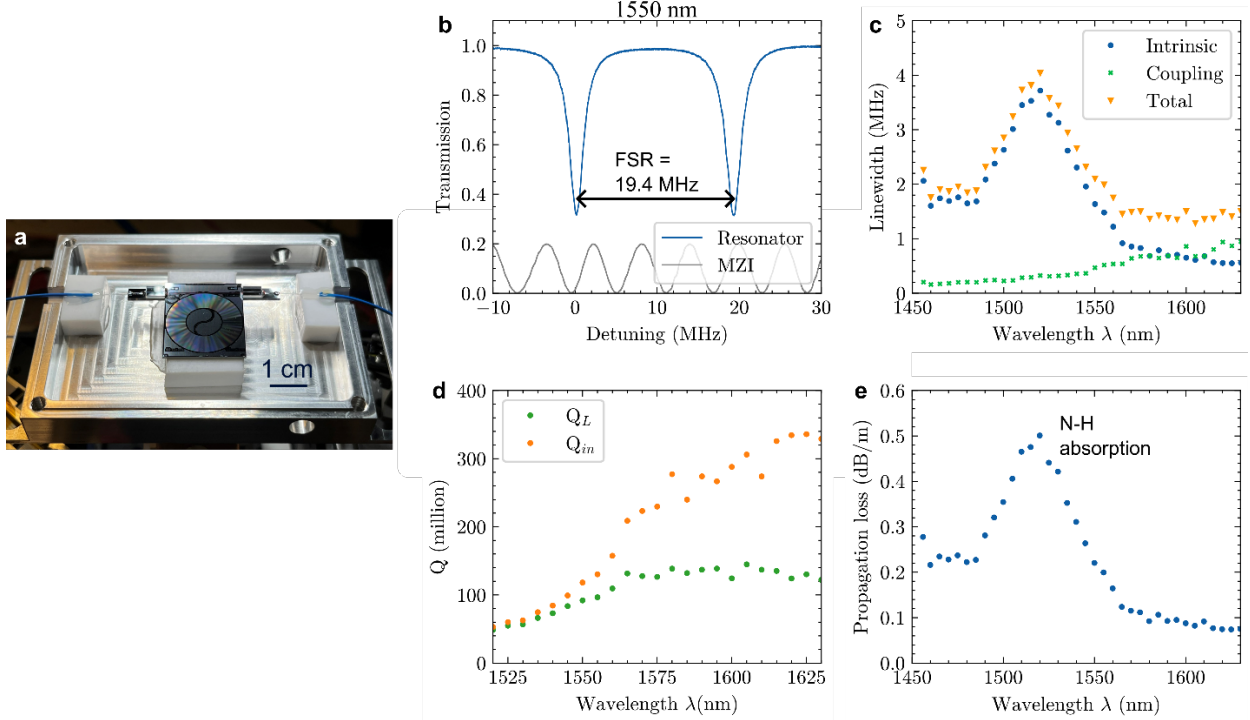

**Supplementary Fig. 2. 10-meter-coil waveguide reference resonator design and testing. a** 10-meter-coil resonator with PM-SMF28 fibers pig-tailed and packaged in a metal enclosure. **b** Spectral scan of the coil resonator resolves the resonance linewidth and FSR. **c** Intrinsic, coupling and total resonator linewidths in MHz are measured from 1450 nm to 1630 nm. **d** Intrinsic and loaded Qs are estimated from the resonator linewidth measurements. **e** Waveguide propagation loss is also estimated, showing an absorption peak around 1520 nm.

## Supplementary Note 4: Frequency noise measurements

Plotted in Supplementary Fig. 3a is the FN of the free running ECTL (blue) measured using an unbalanced fiber-MZI as an optical frequency discriminator (OFD) and reading out the self-delayed homodyne signal on a balanced photodiode. The dashed lines in red, green, and orange indicate calculated estimates of the noise floors of the photodetector noise, thermal refractive noise and photothermal noise of the ECTL intracavity ring resonators. The photo-thermal frequency noise spectrum induced by optical power fluctuations in the external cavity laser is described by,

$$S_{PTN}(f) = \left[\frac{\Delta f_{opt}}{P_{opt}}\right]^2 H_{th}^2(f) P_{opt}^2 S_{RIN} \quad (1)$$

where $S_{RIN}$ is the experimentally measured relative intensity noise (RIN) spectrum (Supplementary Fig. 4c), $P_{opt}$ is the estimated on-chip optical power, $H_{th}(f)$ is the thermal frequency response which can be estimated from COMSOL simulations, and $\frac{\Delta f_{opt}}{P_{opt}}$ describes the photo-thermal red shift strength that includes an estimation of $\xi$ the absorption loss fraction which can be estimated experimentally. Our previous papers provide extensive details on this calculation, for example ref [1].

When we stabilize the ECTL to the 10-meter coil reference cavity the FN at certain low frequency offsets drops by five orders of magnitude, in which case noise due to the fiber-MZI can dominate over the laser noise. To resolve this, we make additional close-to-carrier (CTC) measurements for the locked ECTL by mixing it with a stable reference laser (SRL) system and measuring the heterodyne beatnote signal on a frequency counter. The SRL system includes a Vescent fiber frequency comb locked to a single frequency Rock fiber laser that is itself locked to an ultralow expansion (ULE) cavity. The CTC measurement of the stabilized-ECTL is plotted in Supplementary Fig. 3b (orange) and becomes limited by the speed of the frequency counter at frequency offsets above ~1 kHz. The OFD measurement of the locked-ECTL is plotted (green), and the two can be stitched together (vertical black line) to give a more accurate measurement of the laser FN and to calculate the integral linewidth. See refs [1–5] for more information.

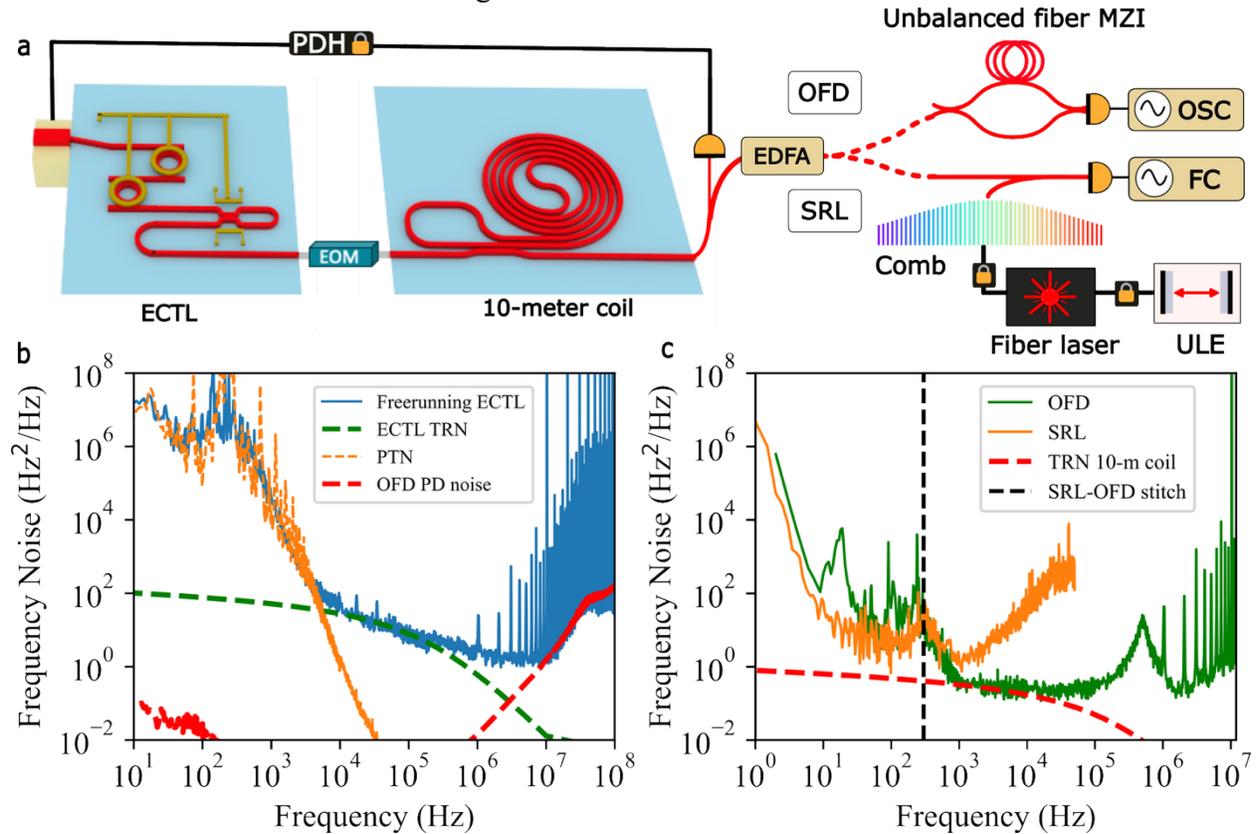

**Supplementary Fig. 3 ECTL frequency noise measurements. a** The frequency noise (FN) of the coil-locked ECTL is measured using two independent techniques: for FN below 3 kHz the ECTL output is mixed with a stable reference laser (SRL) that consists of an optical frequency comb locked to a single frequency Rock fiber laser that in turn is stabilized to an ultralow expansion (ULE) reference cavity, and the heterodyne beatnote signal is measured on a frequency counter (FC). For FN above 3 kHz an unbalanced fiber-MZI is used as an optical frequency discriminator (OFD) and the self-delayed homodyne signal is measured on a balanced photodiode and an oscilloscope (OSC). **b** Frequency noise power spectrum of the freerunning ECTL (blue) at 1550 nm measured using an MZI as an optical frequency discriminator. The orange, red, and green dashed curves are calculated estimates of the thermal refractive and photothermal noise floor of the two intracavity rings and the OFD photodetector noise, respectively. **c** FN of the coil-locked laser measured using two methods: OFD (green) and SRL (orange). The black vertical lines indicates where the two are stitched together for the composite locked-ECTL measurements.

The ECTL operates across 60 nm tuning. FN measurements of the freerunning and locked ECTL

across the tuning range are plotted in Supplementary Fig. 4a,b showing fundamental linewidths between 3-7 Hz and 1/π-integral linewidths from 27 to 60 Hz. Relative intensity noise (RIN) measurements across the ECTL tuning range are also included in Supplementary Fig. 4c.

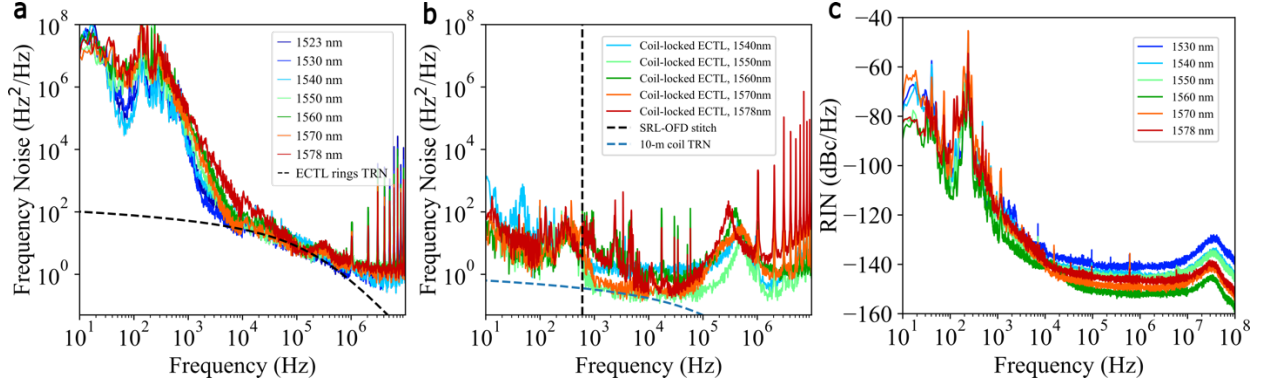

**Supplementary Fig. 4 Additional free running and coil-locked ECTL frequency noise measurements. a** FN measurements of the freerunning ECTL at various points across the tuning range plotted with the calculated TRN limit of the ECTL rings (black-dashed). **b** FN of the ECTL stabilized to the 10-m coil reference cavity at various points across the tuning range using two independent measurement techniques (stitched together at the black vertical line) and plotted with the calculated TRN limit of the 10-m coil (blue-dashed). **c** Measurement of relative intensity noise (RIN) of the ECTL at various points across the tuning range.

## Supplementary Note 5: Optical feedback

Optical feedback in a semiconductor laser can be quantified by the feedback parameter $C$ [6–8],

$$C = \frac{\tau_{ext} r_{ext}}{\tau_{laser} r_{laser}} (1 - |r_{laser}|^2)\sqrt{1+\alpha^2} \quad (1)$$

where $r_{ext}$ and $r_{laser}$ are the reflectivity of the external reflection and the laser mirror, $\tau_{ext}$ and $\tau_{laser}$ are the external cavity and laser cavity round trip lifetimes, and $\alpha$ is the linewidth enhancement factor. We can then compare the relative isolation of two lasers by taking the ratio of their C-parameters where the ratio is only dependent on characteristics of the laser: $\tau, r$ and $\alpha$.

$$\frac{C_1}{C_2} = \frac{\tau_{laser_2}}{\tau_{laser_1}} \frac{r_{laser_2}}{r_{laser_1}} \frac{1-|r_{laser_1}|^2}{1-|r_{laser_2}|^2} \frac{\sqrt{1+\alpha_1^2}}{\sqrt{1+\alpha_2^2}} \quad (2)$$

The robustness of our ECTL to feedback light arises predominantly from two attributes. The first, is the extended cavity photon lifetime, $\tau_{laser}$, due to the two high-Q intracavity rings. We can approximate the overall roundtrip photon lifetime of the dual-ring ECTL as a sum of the Fabry-Perot (FP) and individual ring resonator cavity lifetimes[9]. The FP cavity is formed by the back mirror of the RSOA ($R_1$) and the $Si_3N_4$ waveguide Sagnac loop mirror ($R_2$), and subject to internal losses (η), such as the coupling loss at the interface between the gain chip and the PIC, and has an associated lifetime of:

$$\tau_{FP} = \frac{-2L}{c} ln(R_1 R_2 (1-\eta)^2)^{-1} \quad (3)$$

The rings each contribute an additional $\tau_{RR} = \frac{\lambda Q}{2\pi c}$ and one full roundtrip requires four passes through a ring: $\tau_{ectl} = \tau_{FP} + 4\tau_{RR}$. For the ECTL we estimate that $\tau_{ectl}$ = 2.3 ns. The second attribute that contributes to the robustness to optical feedback is the relatively high reflectivity of the Sagnac loop mirror, which for our design is ~75%. The low loss intracavity rings store enough power within the cavity to support a high mirror reflectivity and still provide useful output power from the laser. Compared to a conventional III-V DFB laser that has an internal Q of ~$1\times10^4$ and front mirror reflectivity of <1 %, we estimate from equation (2) that the ECTL has inherent isolation of approximately 45 dB. Experimentally we demonstrate no degradation in the frequency noise of the ECTL under optical feedback of up to -10 dB. Compared to a commercial III-V DFB that undergoes coherence collapse at -40 dB of feedback, this demonstrates a 30 dB improvement in robustness to feedback.

# Supplementary References

1. Liu, K. *et al.* Photonic circuits for laser stabilization with integrated ultra-high Q and Brillouin laser resonators. *APL Photonics* **7**, 096104 (2022).
2. Liu, K. *et al.* 36 Hz integral linewidth laser based on a photonic integrated 4.0 m coil resonator. *Optica* **9**, 770 (2022).
3. Liu, K. *et al.* Common cavity waveguide coil-resonator stabilized hybrid integrated WDM laser with 89 Hz integral linewidth. in *2024 Optical Fiber Communications Conference and Exhibition (OFC)* 1–3 (2024).
4. Gundavarapu, S. *et al.* Sub-hertz fundamental linewidth photonic integrated Brillouin laser. *Nature Photon* **13**, 60–67 (2019).
5. Chauhan, N. *et al.* Visible light photonic integrated Brillouin laser. *Nat Commun* **12**, 4685 (2021).
6. Zhang, Z. *et al.* High-Speed Coherent Optical Communication With Isolator-Free Heterogeneous Si/III-V Lasers. *Journal of Lightwave Technology* **38**, 6584–6590 (2020).
7. Gomez, S. *et al.* High coherence collapse of a hybrid III–V/Si semiconductor laser with a large quality factor. *J. Phys. Photonics* **2**, 025005 (2020).
8. Harfouche, M. *et al.* Kicking the habit/semiconductor lasers without isolators. *Opt. Express* **28**, 36466 (2020).
9. Oldenbeuving, R. M. *et al.* 25 kHz narrow spectral bandwidth of a wavelength tunable diode laser with a short waveguide-based external cavity. *Laser Phys. Lett.* **10**, 015804 (2012).